\documentclass{article}
\usepackage[utf8]{inputenc}

\usepackage{lscape}

\usepackage{booktabs}

\usepackage{appendix}

\usepackage{amsmath}
\usepackage{amssymb}
\usepackage{parskip}
\usepackage{subfiles}
\usepackage{mathtools}
\usepackage{mathrsfs}
\usepackage{authblk}
\usepackage{soul}

\usepackage{bm}
\usepackage{xcolor}
\usepackage{mystyle}
\usepackage{graphics}
\graphicspath{{./img/}}
\usepackage[position=bottom]{subfig} 

\usepackage{array, booktabs}		

\usepackage[authordate, backend=biber, natbib]{biblatex-chicago}

\bibliography{emh-zotero}

\begin{document}

\title{A Machine Learning approach to Risk Minimisation in Electricity Markets with Coregionalized Sparse Gaussian Processes}
\author[1,2]{Daniel Poh}
\author[1,2]{Stephen Roberts}
\author[1,2*]{Martin Tegn\'{e}r}
\affil[1]{Department of Engineering Science, University of Oxford, Parks Road, Oxford, OX1 2JD, United Kingdom}
\affil[2]{Oxford-Man Institute of Quantitative Finance, Eagle House, Walton Well Road, OX2 6ED, United Kingdom}
\affil[*]{Corresponding author}

\date{March 2019}

\maketitle

\section*{Abstract}

The non-storability of electricity makes it unique among commodity assets, and it is an important driver of its price behaviour in secondary financial markets. The instantaneous and continuous matching of power supply with demand is a key factor explaining its volatility. During periods of high demand, costlier generation capabilities are utilised since electricity cannot be stored and this has the impact of driving prices up very quickly. Furthermore, the non-storability also complicates physical hedging. Owing to these, the problem of joint price-quantity risk in electricity markets is a commonly studied theme. 

We propose using Gaussian Processes (GPs) to tackle this problem since GPs provide a versatile and elegant non-parametric approach for regression and time-series modelling. However, GPs scale poorly with the amount of training data due to a cubic complexity. These considerations suggest that knowledge transfer between price and load is vital for effective hedging, and that a computationally efficient method is required. To this end, we use the coregionalized (or multi-task) sparse GPs which addresses the aforementioned issues.

To gauge the performance of our model, we use an average-load strategy as comparator. The latter is a robust approach commonly used by industry. If the spot and load are uncorrelated and Gaussian, then hedging with the expected load will result in the minimum variance position.

Our main contributions are twofold. Firstly, in developing a coregionalized sparse GP-based approach for hedging. Secondly, in demonstrating that our model-based strategy outperforms the comparator, and can thus be employed for effective hedging in electricity markets.

\textbf{Keywords:} Machine Learning, Gaussian Processes, Energy Risk Management, Electricity Markets

\section{Introduction}
\subsection{Motivation}

The de-regulation of  electricity  markets has led to  increased transparency and the widespread use of risk management products.  On top of playing a vital role balancing demand and supply, power markets fulfil an important function in managing and distributing risk (\citet{bessembinder_equilibrium_2002, botterud_relationship_2010}). Unlike other commodity assets, the non-storability of electricity is a unique feature that has an impact on the term structure since unlike other commodities, the spot and forward prices is no longer linked via the cost of storage.  This feature is thus a key factor not only complicating the fundamentals of hedging and pricing, but also in contributing to volatile spot-markets with structural price jumps (see \citet{botterud_relationship_2010, benth_stochastic_2008} for example).  While market participants hedge with derivatives such as futures or forwards to minimise portfolio risk, the hedging itself may not be executed in an effective fashion. 

We propose a hedging approach that flexibly incorporates the dependency between electricity price and consumption load as well as temporal correlation. We apply our method to data from the UK power market and demonstrate its empirical performance.

\subsection{Related work} \label{sec:lit_rev}

One studied theme within the literature on electricity markets is the coupling of risks arising from quantity and price. This joint risk underscores the importance of modelling the covariation between price and quantity. For instance, \citet{bessembinder_equilibrium_2002} construct an equilibrium-based market model in  which correlation plays a significant part on the optimal hedging strategies in the forward markets. 

\citet{oum_hedging_2006} adopt the position of an agent with access to exogenous spot and forward markets. Using this perspective as a starting point, the authors derive optimal hedging strategies from a utility maximisation approach. These strategies take advantage of  correlation between price and load to manage  joint risk in a single period. \citet{boroumand_hedging_2015} tackle the problem of managing joint risk on an intraday scale. By employing a simulation-based approach, they demonstrate that hedging with shorter time-frequencies can outperform portfolios with long-term focus. 

Close to our work is \citet{tegner_risk-minimisation_2017} who assume the retailer's perspective for risk management, which is a similar position adopted by \citet{oum_hedging_2006}. The former jointly model the electricity spot price and consumption load with a two-dimensional Ornstein-Uhlenbeck process and a seasonality component. 

There are more sophisticated approaches such as the three-factor model of \citet{coulon_model_2013}. This involves load-based regime switching and options that are evaluated with closed form expressions. Here we focus on an operational risk management strategy and use exogenous data from the OTC forward markets.

GPs are not new to electricity markets, although their focus has primarily been on producing single-output forecasts, usually either consumption load or price. 
As mentioned earlier in this section, incorporating correlation between variables is among some of the most important aspects in price and risk models for electricity. The coregionalized or multi-task GP framework provides a natural way of going about this task. Interestingly enough, it has yet to be addressed in the literature on GP-based models applied to the power markets. 

\subsection{Structure}

The remainder of the paper is structured as follows: Section \ref{sec:background} begins with necessary information on the UK electricity market and then goes on to cover the hedging problem as well as a background on GPs. In Section \ref{sec:es} we describe  tools and data used, explain the methodology and experimental setup. Following this, Sections \ref{sec:gp_comp} and \ref{sec:hedging_results} present the results of the model construction and the hedging experiment respectively. Finally, Section \ref{sec:conclusion} recaps and summarises key insights, and highlight possible directions for future work.

\section{Background}
\label{sec:background}
\subsection{The UK electricity sector}
The power market in the UK can be broadly thought of as having two layers: the wholesale market and the retail market. In the wholesale market, power is produced by generators and sold to suppliers (or retailers). The retail market is the next layer where suppliers then resell electricity to end-users.

The wholesale electricity market can be further divided into three ‘sub-markets’ that have a temporal ordering. The future and forward markets typically occupy the longer end of this spectrum, dealing with deliveries over a longer span of time such as a month or a quarter. These contracts obligate participants to either consume or deliver a fixed amount of electricity during a specific time period in the future based on some mutually agreed price at inception. The day-ahead market has a shorter horizon. Electricity in this market is transacted one day prior to actual delivery. Additionally, the term ‘spot-price’ used in most electricity markets (in addition to the UK's) refers to the price in the day-ahead market. The intra-day market and/or the balancing mechanism has an even shorter time horizon. Electricity traded on these markets are delivered on the same day itself. Trading activities within these sub-markets usually take place either over-the-counter (OTC) or on exchanges. 

\subsection{Hedging problem}

While there exist numerous financial products that can be used to manage power risk, we consider only the use of two in this paper: the base load and the peak load forward contract for UK power. Both are OTC instruments that require settling the price difference between spot and some agreed amount, over some period of time (a month in our context). While the base contract pays this difference for every hour of the month, the peak contract applies only for “peak hours”, which is from 7 a.m. to 7 p.m. on weekdays only. By constructing a portfolio of base and peak load contracts, we are able to create a portfolio that resembles the load profile.

The setup of the following hedging problem where the aim is to accurately replicate an uncertain future financial obligation is similar to \citet{tegner_risk-minimisation_2017}. This is usually achieved by using financial derivatives such as futures and forwards. Specifically, we attempt to minimize the expected loss by determining the portfolio of base and peak load forwards to hold.

To this end, let the (positive) payoff of the hedged portfolio at time $T$ be $\pi_T$. Thus, the loss is given by $-\pi_T$. Additionally, for some loss function $u(\cdot)$, a measure of its risk (See \citet{artzner_coherent_1999} for more details of risk measures) is given by $ \mathbb{E}[u(-\pi_T)]$, where $\mathbb{E}(\cdot)$ is the expectation operator. There are many options for picking the loss metric $u(\cdot)$ and some of these are the maximum loss, quadratic loss and the ‘hockey-stick’ loss (also known as the rectified linear unit). In order to penalize losses more, we use the exponential loss $u(\cdot)=\exp(\cdot)$.

Suppose that at time $t_0$, we enter an agreement to deliver at some future time $T>t_0$ an unspecified amount $L_T$ of some commodity at a (constant) price $C$ contracted by a fixed price agreement. If we take a naked position, this would mean having to purchase the commodity from the market at the prevailing price $S_T$ and volume $L_T$. The payoff at time $T$ is thus 
\begin{equation}
    \pi_T = (C - S_T)L_T \nonumber
\end{equation}

In order to reduce the exposure to price fluctuations, we initiate a static hedge at $t_0$. This can be achieved by purchasing a forward contract at price $F$ with expiry $T$, although the volume $V$ needs to be established at the time of purchase. If the fixed price agreement differs from the futures by some margin $\delta \geq 0$ such that $C = F + \delta$, then the hedged position has payoff 
\begin{equation}
    \pi_T = (S_T - F)(V - L_T) + \delta L_T   \label{eqn:hedged_pos}
\end{equation}

where we would need to determine the optimal $V$ at the point when we set up our hedge. For this purpose, we use the exponential loss function and minimize the expected loss with respect to $V$. In other words, we want to find the optimal $V^*$ that minimizes 
\begin{equation}
    f(V) = \mathbb{E} \Bigl [ u \Bigl ( - (S_T - F)(V - L_T) + \delta L_T \Bigr ) \Bigr ] \nonumber
\end{equation}

Bringing the UK power markets back in to focus, suppose we are contracted to deliver based on the terms of some fixed price agreement for the month $M$. We denote with $T_i$ the $i^{\textrm{th}}$ hour from the point of initiating the hedge to delivery. Additionally, for a given month $M$ we refer to the set of  peak hours as $M_p$, and the set of off-peak hours as $M_o$. Hence from equation (\ref{eqn:hedged_pos}), we can write the off-peak payoff for $T_i \in M_o$ as
\begin{align}
    \pi_{T_i} & = (S_{T_i} - F^b)(V^b - L_{T_i}) + \delta^b L_T \nonumber \\
    & = (S_{T_i} - F^b)(V^b - L_{T_i}), \quad T_i \in M_o   \label{eqn:off_peak_payoff}
\end{align} 

where $V^b$  is understood to be the base load forward position with price $F^b$. For convenience, we set the differential between the fixed price agreement and base load forward $\delta^b$ to 0.

To obtain the payoff for a peak hour, we would combine the peak load forward position $V^p$ and price $F^p$ with a base load forward position. It is usually safe to assume that $F^p \geq F^b$. The peak payoff for a hedged position is then
\begin{align}
    \pi_{T_i} = & \; (S_{T_i} - F^b)V^b + (S_{T_i} - F^p)V^p + \nonumber \\ 
                & (F^{\textrm{p-fpa}}-S_{T_i})L_{T_i} , \quad T_i \in M_p \nonumber
\end{align}
where, similar to the off-peak case, we set the margin $\delta^p = 0$. If we define 
\begin{equation}
    \tilde{F} \equiv F^p - \dfrac{V^b}{V^b + V^p} (F^p-F^b) \nonumber
\end{equation}
such that $F^b \leq \tilde{F} \leq F^p$, we can compact the expression for the peak payoffs as
\begin{equation}
    \pi_{T_i} = (S_{T_i} - \tilde{F})(V^b + V^P - L_{T_i}) \nonumber
\end{equation}
where $T_i \in M_p$.

If the goal is to hedge in an optimal fashion for the month $M$, this means having to determine $V^{b*}$ and $V^{p*}$ that minimizes the combined loss function
\begin{align}
    f(V^b, V^p)  = & \; \sum_{T_i \in M_o} \mathbb{E} \Bigl [ u \Bigl (-(S_{T_i}-F^b)(V^b - L_{T_i}) \Bigr ) \Bigr ] \nonumber \\ 
    & + \sum_{T_i \in M_p} \mathbb{E} \Bigl [ u \Bigl (-(S_{T_i}-\Tilde{F})(V^b + V^p - L_{T_i}) \Bigr ) \Bigr ] \label{eqn:loss_func_fv}
\end{align}

where, similar to equation (\ref{eqn:off_peak_payoff}), both differentials $\delta^b$ and $\delta^p$ of the respective forwards to the fixed price agreement are set to 0 for convenience. A terser formulation of the above problem is 
\begin{equation}
    \argmin_{V^b, V^p} f(V^b, V^p) \nonumber
\end{equation}
There are two approaches to solve this minimization problem: We can either determine an expression for the expectation given by (\ref{eqn:loss_func_fv}) where $\delta L_{T_i} = 0$ for simplicity, or we can arrive at an approximate value via Monte-Carlo simulation. We adopt the latter approach.

\subsection{Gaussian Processes}

Gaussian processes are an extension of multivariate Gaussian distributions. We briefly go through the underlying fundamentals in this section. A detailed treatment can be found in \citet{rasmussen_gaussian_2006}. 

A GP defines a probability distribution over functions. For a given input space $\mathcal{X}$, a GP is defined by a mean function $m(\bm x)$ and a covariance function $\kappa(\bm x, \bm x')$ as
\begin{equation}
    f(\bm x) \sim \mathcal{GP}(m(\bm x), \kappa(\bm x, \bm x')) \nonumber
\end{equation}
where
\begin{align}
    m(\bm x) & = \mathbb{E}[f(\bm x)] \nonumber \\
    \kappa(\bm x, \bm x') & = \mathbb{E} \Bigl [ \Bigl ( f(\bm x) - m(\bm x) \Bigr ) \Bigl ( f(\bm x') - m(\bm x') \Bigr )^\top \Bigr ] \nonumber
\end{align}
The mean function $m(\bm x)$ is often set to zero since the GP is flexible enough to model the mean arbitrarily well. 

Suppose we now have a training dataset $(\bm X, \bm y)$ as well as a test dataset $(\bm X_*, \bm y_*)$, where $\bm X\in \mathbb{R}^{N \times D}$, $\bm y \in \mathbb{R}^N$, $\bm X_* \in \mathbb{R}^{N* \times D}$ and $\bm y_* \in \mathbb{R}^{N*}$. Denoting the function outputs of the training and test data by $\bm f(\bm X)$ and $\bm f(\bm X_*)$, which we shorten to $\bm f$ and $\bm f_*$ for brevity, we can make use of the training observations to make predictions on the test set by the following joint distribution
\begin{equation}
    \begin{pmatrix}
        \bm f \\
        \bm f_*
    \end{pmatrix}
        \sim
    N
    \begin{pmatrix}
        \bm 0
        ,
        \begin{pmatrix}
            \bm K (\bm X, \bm X) & \bm K (\bm X, \bm X_*) \\
            \bm K (\bm X_*, \bm X) & \bm K (\bm X_*, \bm X_*)
        \end{pmatrix}
    \end{pmatrix}
    \label{eqn:gp_joint_distrib}
\end{equation}
where $\bm K (\bm X, \bm X)$ is a covariance matrix where the $(i,j)^\textrm{th}$ element is is $\kappa(\bm x_i, \bm x_j)$. The posterior then has the form
\begin{align}
    p(\bm f_* | \bm X_*, \bm X, \bm f)    & = N(\bm f_* | \bm \mu_* , \bm \Sigma_*) \nonumber \\
    \bm \mu_*                             & = \bm K^\top_* \bm K^{-1} \bm f  \label{eqn:post_mean_noisefree} \\
    \bm \Sigma_*                          & = \bm K_{**}  - \bm K^\top_* \bm K^{-1} \bm K_* \label{eqn:post_covar_noisefree}
\end{align}

\subsubsection{Covariance Functions}

The GP's covariance functions can be used to encode prior domain knowledge about $f$. Intuitively, these functions allow for generalization of the model by correlating new inputs to existing observations.

For this subsection, we let $r \equiv |\bm x- \bm x'|$. The most commonly used covariance function is the squared exponential (SE) or radial basis function, which has the form

\begin{equation}
    \kappa_{\textrm{SE}}(r) = \sigma^2 \exp \Bigl ( -\dfrac{r^2}{\ell^2} \Bigr )
    \nonumber
\end{equation}
The parameters $\sigma$ and $\ell$ control the amplitude and characteristic length scale respectively. The continuous, differentiable and stationary properties of the SE kernel make it a popular choice for generic modelling. 

As the SE is infinitely differentiable, it yields smooth sample paths which might be unsuitable for real-world phenomena. An alternative to the SE kernel is the Mat\'{e}rn family, for which the general form is
\begin{equation}
    \kappa_{\text{Mat\'{e}rn}}(r) = \dfrac{2^{1-\nu}}{\Gamma(\nu)} \Bigl ( \dfrac{\sqrt{2\nu}r}{\ell} \Bigr )^\nu K_\nu \Bigl ( \dfrac{\sqrt{2\nu}r}{\ell} \Bigr )  \nonumber
\end{equation}
where $\sigma$ and $\ell$ are both positive parameters, while $K_\nu$ is a modified Bessel function. The form for the Mat\'{e}rn class simplifies  if it is half integer, that is, $\nu=p+1/2$  where p is a non-negative integer. For most machine learning applications, values of $\nu=3/2$ and $\nu=5/2$ are commonly encountered. We make use of the latter configuration in this paper. 

The periodic covariance function is used to model functions that are associated with some characteristic periodicity. It has the form
\begin{equation}
    \kappa_{\textrm{Periodic}}(r) = \sigma^2 \exp \Bigl (-\dfrac{2\sin^2(\pi(r)/p)}{\ell^2} \Bigr ) \nonumber
\end{equation}
where $\sigma$ , $p$ and $\ell$ are parameters for the amplitude, period and length scale respectively. It is a well-known that electricity price and load exhibit periodicity on multiple levels. As we explain later, a combination of these kernels parameterized with different periods will be used to capture these seasonalities.
The rational quadratic is the last covariance considered for our model. It has the form 
\begin{equation}
    \kappa_{\textrm{RQ}}(r) =  \sigma \Bigl ( 1 + \dfrac{r^2}{2\alpha \ell^2} \Bigr )^{-\alpha} \nonumber
\end{equation}
where the parameters $\sigma$, $\alpha$ and $\ell$ are respectively the amplitude, shape parameter and length scale. The shape parameter determines the diffuseness of the length scales. The rational quadratic is generally used to model small to medium term irregularities.

One way of constructing new kernels is by affine transformations. We restrict the scope of our study to composite kernels formed by addition. A wider list of stationary and non-stationary covariance functions can be found in \citet{rasmussen_gaussian_2006}.

\subsection{Sparse Gaussian Processes}
\label{sec:bkg_sparse_gps}

GPs are flexible but perform poorly on larger data sets due to the matrix inversion operation (see Equations (\ref{eqn:post_mean_noisefree}) and (\ref{eqn:post_covar_noisefree})) which scales as $O(n^3)$. This limitation has motivated work in various computationally efficient approaches that aim to  approximate the precise GP solution (\citet{gal_distributed_2014, hensman_gaussian_2013}). The comprehensive review by \citet{liu_when_2018} classifies scalable GPs by first grouping them into those that produce global approximations and those that produce local approximations. Methods falling within global approximations can be further sub-divided into those that (i) operate on a subset of the training data, (ii) use sparse kernels, and (iii) employ sparse approximations. Our GP model uses a variant of the latter known as the Deterministic Training Conditional (DTC). The rest of this section outlines the general idea of  sparse approximations, and readers are encouraged to refer to  \citet{quinonero-candela_unifying_2005} for details.

We start by modifying the joint distribution given by Equation (\ref{eqn:gp_joint_distrib}) to reduce the computational load due to matrix inversion in the posterior distribution. This step involves introducing latent or inducing variables $\bm u = (u_1, u_2, \hdots ,u_m)^\top$. These inducing variables correspond to a set of input locations, hence they are also known as inducing inputs. Sparse algorithms vary in their approach of selecting inducing variables.
By the consistency property of GPs, we can recover $p(\bm f, \bm f_*)$ from $p(\bm f, \bm f_*, \bm u)$ by integrating out $\bm u$ in the latter
\begin{equation}
    p(\bm f, \bm f_*) = \int p(\bm f, \bm f_* | \bm u) p(\bm u) \, d \bm u \nonumber
\end{equation}
where $\bm u \sim N(\bm 0, \bm {K_{u,u}})$. By assuming that both $\bm f$ and $\bm f_*$ are conditionally independent given $\bm u$, the joint of the prior is approximated as
\begin{align}
    p(\bm f, \bm f_*) 
    & \approx q(\bm f, \bm f_*) \nonumber \\
    & = \int q(\bm f | \bm u) q( \bm f_* | \bm u) p(\bm u) \, d \bm u \nonumber 
\end{align}
This approximation is the basis upon which many sparse approximation techniques are built. Additionally, different assumptions on the approximate training conditional $q(\bm f| \bm u)$ and approximate test conditional $q(\bm f_*| \bm u)$ give rise to different algorithms. The exact train and test conditionals are given respectively as
\begin{align}
    p(\bm f | \bm u) & = N \Bigl ( \bm K_{\bm f, \bm u}\bm K_{\bm u,\bm u}^{-1} \bm u, \, \bm K_{\bm u,\bm u} - \bm Q_{\bm f,\bm f} \Bigr ) \nonumber \\
    p(\bm f_* | \bm u) & = N \Bigl ( \bm K_{\bm f_* , \bm u}\bm K_{\bm u, \bm u}^{-1} \bm u, \, \bm K_{\bm f_*, \bm f_*} - \bm Q_{\bm f_*,\bm f_*} \Bigr ) \nonumber
\end{align}
where $\bm{Q}_{\bm a, \bm b} \equiv  \bm K_{\bm a , \bm u} \bm K^{-1}_{\bm u, \bm u} \bm K_{\bm u, \bm b}$.

Another approach to sparse approximation by \citet{seeger_fast_2003} makes use of an estimation of the likelihood via the projection $\bm f = \bm K_{\bm f, \bm u} \bm K^{-1}_{\bm u, \bm u} \bm u$ which gives
\begin{align}
    p(\bm y | \bm f) 
    & \approx q(\bm y | \bm u) \nonumber \\
    & = N \Bigl ( \bm K_{\bm f, \bm u} \bm K^{-1}_{\bm u, \bm u} \bm u, \, \sigma^2_{\textrm{noise}} \bm I \Bigr ) \nonumber
\end{align}
The DTC achieves an equivalent model but makes use of a deterministic training conditional and exact test conditional which are given as
\begin{align}
    q_{\textrm{DTC}}(\bm f | \bm u) & = N \Bigl ( \bm K_{\bm f, \bm u} \bm K^{-1}_{\bm u, \bm u} \bm u, \, \bm 0 \Bigr ) \nonumber \\
    q_{\textrm{DTC}}(\bm f_* | \bm u) & = p(\bm f_* | \bm u) \nonumber
\end{align}
The posterior or predictive distribution under the DTC is

\begin{equation}
    \begin{split}
        q_{\textrm{DTC}} 
    & = N \Bigl (\bm Q_{\bm f_*, \bm f}( \bm Q_{\bm f, \bm f} + \sigma^2_{\textrm{noise}} \bm I)^{-1} \bm y, \\
    &\qquad \qquad \bm K_{\bm f_*, \bm f_*} - \bm Q_{\bm f_*, \bm f} ( \bm Q_{\bm f, \bm f} + \sigma^2_{\textrm{noise}} \bm I  \Bigr)^{-1} \bm Q_{\bm f, \bm  f_*} \Bigr ) \\
    & = N \Bigl ( \sigma^{-2}_{\textrm{noise}} \bm K_{\bm f_*, \bm u} \bm \psi \bm K_{\bm u, \bm f} \bm y, \\
    &\qquad \qquad \bm K_{\bm f_*, \bm f_*} - \bm Q_{\bm f_*, \bm f_*} + \bm K_{\bm f_*, \bm u} \bm \psi \bm K^\top_{\bm f_*, \bm u} \Bigr ) \nonumber
    \end{split}
\end{equation}
where $\bm \psi \equiv \Bigl ( \sigma^{-2}_{\textrm{noise}} \bm K_{\bm u, \bm f} \bm K_{\bm f, \bm u} + \bm K_{\bm u, \bm u} \Bigr )^{-1}$.

\subsection{Coregionalized Gaussian Proceses}

Coregionalized GPs, also known as multi-task GPs, essentially extend the concept of correlating data to GPs. It suggests that the information gained from one process can be generalized to another; in other words, knowledge is transferred from one process to another. We cover the main ideas underlying coregionalization in this section, further details can be found in \citet{alvarez_kernels_2011}.

We can motivate this idea by first supposing a set of models, $\mathcal{M}$. Let the corresponding dataset used by some model $m \in \mathcal{M}$ be denoted by the scalar vector $\bm x$ containing $P$ elements. For simplicity, assume that all datasets are of size $P$. We can then denote a model and its dataset by a tuple $(\bm x, m)$. To allow knowledge transfer among models, we introduce some covariance kernel $\mathcal{K}$ that describes the correlation between the models $m$ and $m'$ using some matrix $B_{m,m'}$. This can be formulated as
\begin{align}
    \mathcal{K}((\bm x, m), (\bm x', m')) = B_{m,m'} \times K(\bm x, \bm x') \label{eqn:cgp_1}
\end{align}
where it is understood that $\mathcal{K}$ also encodes the parameters of $K(\cdot, \cdot)$. The kernel matrix corresponding to ($\ref{eqn:cgp_1}$) can be written as
\begin{align}
    \bm K( \bm x, \bm x') 
        & = 
        \begin{pmatrix}
        B_{1,1} \times K(\bm x, \bm x') & \hdots & B_{1,D} \times K(\bm x, \bm x') \\
        \vdots & \ddots & \vdots \\
        B_{D,1} \times K(\bm x, \bm x') & \hdots & B_{D,D} \times K(\bm x, \bm x') 
        \end{pmatrix} \nonumber \\
        & = \bm B \otimes K (\bm x, \bm x') \nonumber
\end{align}
where $D$ is the number of elements in $\mathcal{M}$, and $\bm B \in \mathbb{R}^{D \times D}$ is a coregionalization matrix. $\bm K$ is $DP \times DP$.

In order that the multiple output kernel $\bm K$ qualifies as a valid kernel, we require that both $K$ and $\bm B$ are valid covariance matrices. However, if $K$ is already valid, then we only require that $\bm B$ be positive definite. 

\section{Empirical study}\label{sec:es}
This section details a hedging approach with the GP model. We first introduce and describe the datasets and the tools used. We then describe how we construct the kernel. Following that, we explain how model estimation is carried out and conclude by detailing the setup of the hedging problem.

\subsection{Data and implementation tools}

Hourly datasets for the UK day-ahead electricity spot price are obtained from \citet{nord_pool_n2ex_2018}. Prices for the OTC base and peak load forward contracts are obtained from \citet{bloomberg_uk_nodate}. 

To the best of our knowledge, there is no publicly available hourly consumption load data for the UK from 2016 to 2018. We worked around this issue by estimating consumption load from power demand data sourced from the \citet{eso_nodate}. In order to do this, we use the facts that load is a flow of power over some period while demand is a snapshot at a single point in time with units of measurements MWh (Megawatt hour) and MW (Megawatt) respectively. The available demand data from National Grid ESO are snapshots recorded at thirty minute intervals. By averaging two thirty-minute readings (with the first starting exactly on the hour) and then assuming that this mean is constant over the hour, we obtain an approximate measure of consumption load.

\subsection{Exploratory data analysis and pre-processing}\label{sec:es_eda}

\begin{table}[t!]  

	\setlength\extrarowheight{7pt}
	\fontsize{8}{11}\selectfont
	\centering
	\renewcommand{\arraystretch}{0.5}
	\scalebox{0.75}{
	\begin{tabular}{m{1.4in} m{.6in} m{.6in} m{.6in} m{.6in} m{.6in} m{.6in}}
		\toprule
                              & SD \textless 1 & 1 \textless SD \textless 2 & 2 \textless SD \textless 3 & 3 \textless SD \textless 4 & 4 \textless SD \textless 5 & SD \textgreater 5 \\
		\midrule
Off-peak price                & 13190          & 1121                        & 53                         & 11                      & 7                          & 13                \\ 
Off-peak price (\%)           & 91.63          & 7.79                       & 0.37                       & 0.08                       & 0.05                       & 0.09              \\ 
Peak price                    & 12755          & 1484                       & 116                        & 16                         & 4                          & 16                \\ 
Peak price (\%)               & 88.61          & 10.31                       & 0.81                       & 0.11                       & 0.03                       & 0.11              \\ 
\midrule
Norm. Off-peak load      & 9455           & 4455                       & 485                        & 0                          & 0                          & 0                 \\ 
Norm. Off-peak load (\%) & 65.68          & 30.95                      & 3.37                       & 0.00                       & 0.00                       & 0.00              \\ 
Norm. Peak load          & 9753           & 3917                       & 658                        & 63                         & 0                          & 0                 \\ 
Norm. Peak load (\%)     & 67.75          & 27.22                      & 4.57                       & 0.44                       & 0.00                       & 0.00       \\
		\bottomrule
	\end{tabular}
	}
	\caption{Distribution of next day spot-price and power load (19 September 2015 to 31 December 2018)}
    \label{tbl:nd_spot_dist}
\end{table}

Both electricity demand and consumption load within the UK exhibit repetitive behaviors on multiple time-scales: on a yearly/seasonal basis, across the week and over a day (\citet{gavin_seasonal_2014}). While price generally moves in tandem with load, its jumps makes it far more volatile. This is evident from Table. \ref{tbl:nd_spot_dist}\footnote{The window starts from 19 September 2015 because the some of the hedging models for January 2016 (i.e. the CSGP-3M variants) were trained on 3 months' of earlier data.}, where we can see a number of peak and off-peak prices going further than three standard deviations from their respective means. While there also appears to be a fair number of load points between two and three deviations, this should be expected given the seasonality of the data.

For data pre-processing, we smooth the spikes in price by setting them to be no more than three standard deviations from the mean. When this condition is met for price, we also apply this operation to load. This is an important step to ensure that the optimization produce posteriors that reasonably fit the data. An alternative would have been to fix the length scale parameter in each of the covariance functions, although this would be both cumbersome and less intuitive.

\subsection{Kernel construction}\label{sec:es_kern_cons}
We assembled the composite kernel by summing four types of kernels: the square exponential (SE), Mat\'{e}rn with $\nu=5/2$, periodic and rational quadratic kernels. The SE is included to capture the broader trend underlying the data sets, while the  Mat\'{e}rn kernel incorporates the non-smooth nature of both price and load. We model the repetitive nature of the dataset over various time frequencies with three separate periodic kernels with periods of 12, 24 and 168 hours. These settings are based on domain knowledge and are confirmed in the plots and discussions in Section \ref{sec:es_eda}. Finally, the rational quadratic kernel is added to model the noise term in the datasets. Taking these together, the final composite kernel is
\begin{equation}
    \kappa_{\textrm{composite}} = \kappa_{\textrm{SE}} + \kappa_{\textrm{Mat52}} + \kappa_{\textrm{Per12}} + \kappa_{\textrm{Per24}} + \kappa_{\textrm{Per168}} + \kappa_{\textrm{RQ}}
    \label{eqn:final_comp_kern}
\end{equation}
We restrict ourselves to adding simple kernel components to maintain a high degree of explainability. While not explicit, note that a white noise kernel is added to Equation (\ref{eqn:final_comp_kern}) to account for observations variance.

\subsection{Hyper-parameter tuning and model search}
We define our primary model to be the coregionalized sparse GP (CSGP) with kernel given by (\ref{eqn:final_comp_kern}) trained on 30 days' (approximately 1 month) of hourly data with 10\% sparsity (10\% of training data). While we recognize that sparsity could be a tunable hyper-parameter,  we fix it here to control the computational complexity of the algorithm. 
 We tune the hyper-parameters for the respective covariance functions listed in Section \ref{sec:bkg_sparse_gps}. Note that there are only six such variables in total for the three periodic functions since we have fixed  periodicities. In order to find the optimal set of parameters for each hedging month, we run the optimization a few times to avoid running into local optima. We do this for each month.

\subsection{Hedging problem setup}
For a particular hedging month, the portfolio delivering the optimal hedge with respect to our model is
\begin{align}
    & \argmin_{V^b, V^p} f(V^{b}, V^{p} | \bm \theta) \nonumber \quad \textrm{s.t.} \quad 0 < F^b \leq \tilde{F} \leq F^p \nonumber \\
    & \;\textrm{where} \quad \tilde{F} \equiv F^p - \dfrac{V^b}{V^b + V^p} (F^p-F^b)
\end{align}
where $\bm \theta$ is the vector of optimized parameters of the fitted CSGP, while $F^b$ and $F^p$ are the prices of the peak load and base load forward contract for that month.

To hedge a given month, we purchase some combination of base and peak load contracts around two weeks before the start of the month for liquidity considerations. Therefore, to hedge an exposure for the whole of January in 2018, we would buy the appropriate amount of contracts on the 18th December 2018 with the model trained on the hourly data from the previous thirty days. We assume that the base and peak load forward contracts are purchased at the closing price on the hedging initialization date. To ensure numerical stability, each hourly load data for the training month is rebased against the maximum load for the length of entire study. This has the effect of converting the absolute optimum base and peak positions to percentages of the overall maximum load. The actual payoff for $T_i$ or for the $i^{\textrm{th}}$ hour is given as

\[
  \pi_{t_i} = \left.
  \begin{cases}
    (S_{t_i} - F^b) ( V^{b*} - L_{t_i})                 & \quad \for t_i \in M_o \\
    (S_{t_i} - \tilde{F}) ( V^{b*} + V^{p*}- L_{t_i})   & \quad \for t_i \in M_p
  \end{cases}
  \right.
\]

\begin{align}
    \pi_{t_i} & = (S_{t_i} - F^b) ( V^{b*} - L_{t_i}) \quad \for t_i \in M_o \\
    \pi_{t_i} & = (S_{t_i} - \tilde{F}) ( V^{b*} + V^{p*}- L_{t_i}) \quad \for t_i \in M_P
\end{align}

The table in Appendix \ref{appen:A} lists the various dates at which we initiated our monthly positions for the entire period of our study.

\section{Comparing performance across GP configurations}
\label{sec:gp_comp}
In the following set of studies, we refer to CSGPs trained on one month of data as the CSGP-1M, on two months as CSGP-2M, and so on. We compare our primary model, which is the CSGP-1M on 10\% sparsity using the full kernel given by Equation (\ref{eqn:final_comp_kern}), in three different settings.

\subsection{Coregionalized GPs using the full kernel and trained on one month of data with different sparsity levels}
\label{sec_gp_comp_exp_1}

\begin{figure}[th!]
	\centering
	\subfloat[Price forecast with 10\% of 1M data]{
		\includegraphics[width=0.49\textwidth] {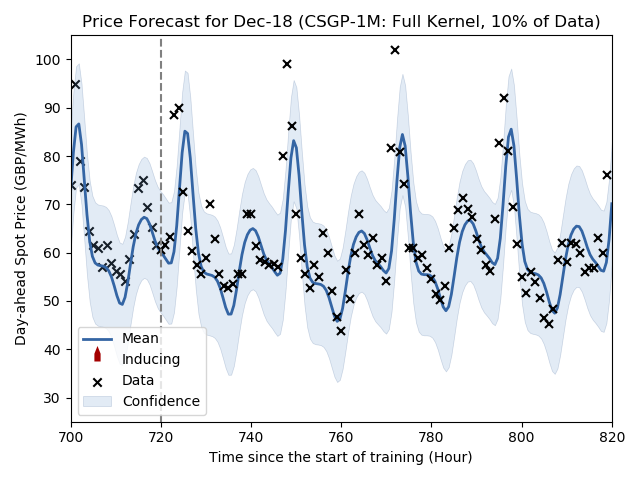}
		\label{fig:csgp_1m_sp010_price}
		} 
	\hspace*{-1em}
	\subfloat[Load forecast with 10\% of 1M data]{
		\includegraphics[width=0.49\textwidth] {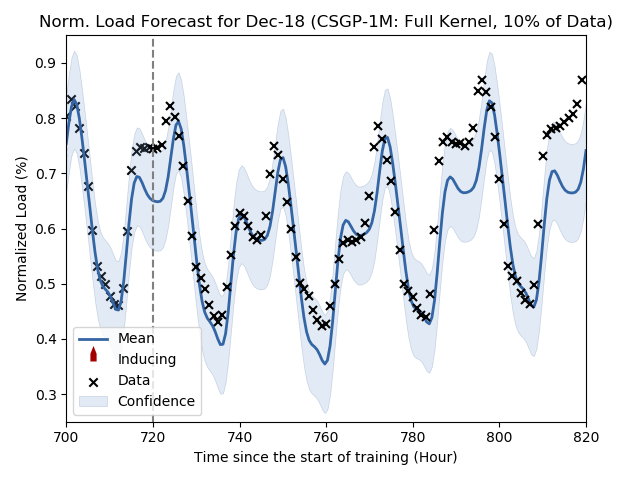}
		} \\
	\subfloat[Price forecast with 1\% of 1M data]{
		\includegraphics[width=0.49\textwidth] {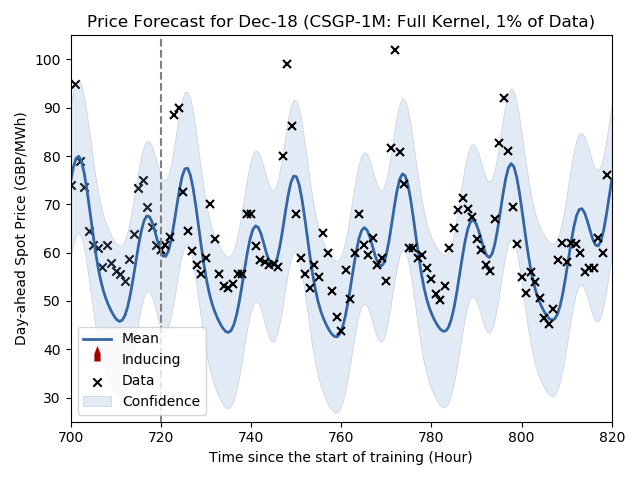}
		\label{fig:csgp_1m_sp001_price}
		} 
	\hspace*{-1em}
	\subfloat[Load forecast with 1\% of 1M data]{
		\includegraphics[width=0.49\textwidth] {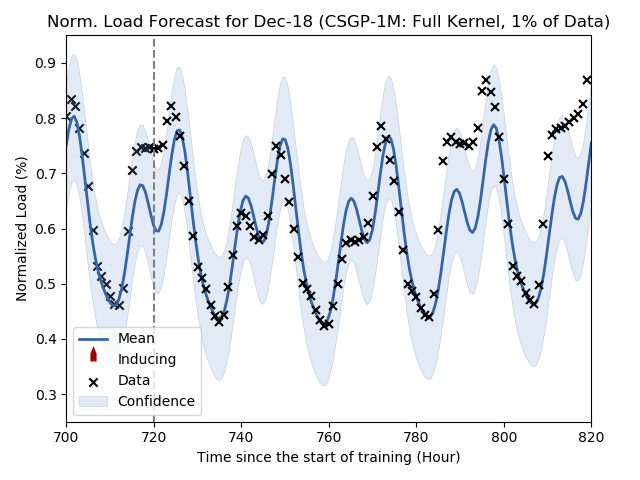}
		} \\
	\subfloat[Price forecast with 100\% of 1M data]{
		\includegraphics[width=0.49\textwidth] {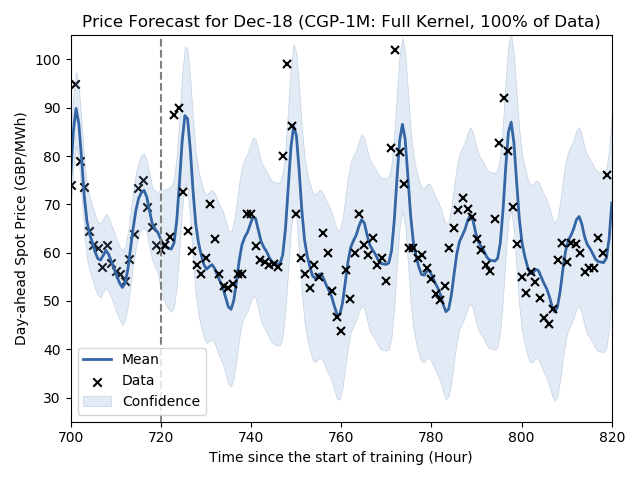}
		\label{fig:cgp_1m_price}
		} 
	\hspace*{-1em}
	\subfloat[Load forecast with 100\% of 1M data]{
		\includegraphics[width=0.49\textwidth] {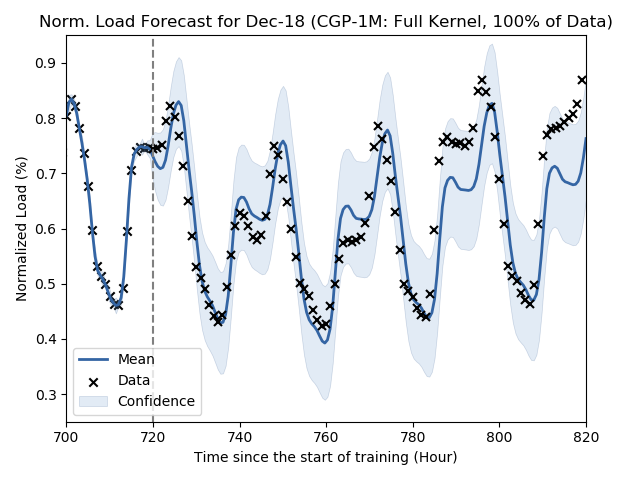}
		} \\
	\caption{Examining the resulting price and load forecasts for December 2018 arising from models trained on 1M of data with different degrees of sparsity. The top pair shows the posterior on spot price (left) and consumption load (right) using a sparsity of 10\% sparsity. The middle pair uses 1\% sparsity while the bottom pair is trained on the full data set and does not use sparsity at all.
	}
	\label{fig:exp_1}
\end{figure}

Fig. \ref{fig:exp_1} shows the posterior predictions on price and load at different degrees of sparsity while fixing all other features of the model.  These models are trained on 720 hours (one month) of hourly data for hedging on December 2018, and prediction takes place from the $720^{\text{th}}$ hour onwards, i.e., 1st December 2018 midnight onwards. The dashed vertical line marks the point beyond which we start generating forecasts for the hedging month of December itself. 

As seen in Fig. \ref{fig:exp_1}, the posterior mean (the dark blue line) becomes less smooth as more points are used for training\footnote{To clarify the use of \textit{sparsity} in the context of this paper, for two models trained on the same data, the model with 1\% sparsity uses less data than the model using say, 10\% of the data}. Using fewer points and ensuring that they are sufficiently far apart from one another produces longer correlation lengths between points. This allows us to learn a stable longer term trend while avoiding over-fitting to the short term noise contained in the data. For example, the mean does not attempt to fit the price spikes in Fig. \ref{fig:csgp_1m_sp001_price} (using 1\% of training data) unlike Fig. \ref{fig:cgp_1m_price} (100\% of data). This has important implications for hedging performance as we shall see later in the paper.

\subsection{CSGP-1M fixed at 10\% sparsity with different kernel components removed}
We find that the set of periodic functions are the largest contributor to forecast performance. Models incorporating these components but lack any of the other components in Equation (\ref{eqn:final_comp_kern}) generates predictions that are similar to the base model. Among these models, it is difficult to definitively say which is better since their differences are marginal.

\subsection{CSGP fixed at 10\% sparsity using the full kernel but trained on different lengths of time}

\begin{figure}[th!]
	\centering
	\subfloat[Price forecast with 10\% of 2M data]{
		\includegraphics[width=0.49\textwidth] {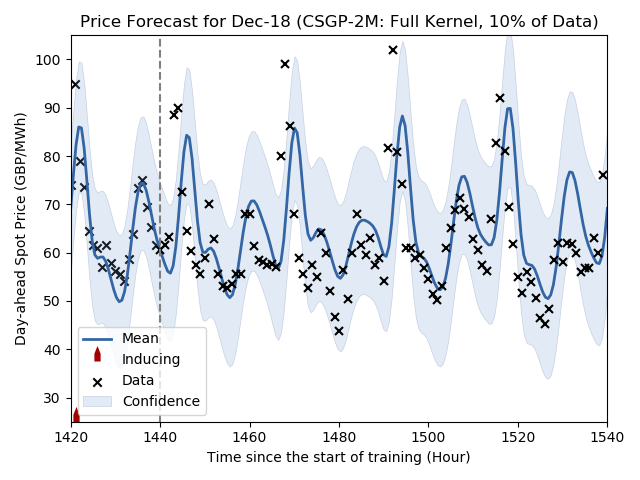}
		\label{fig:csgp_2m_price}
		} 
	\hspace*{-1em}
	\subfloat[Load forecast with 10\% of 2M data]{
		\includegraphics[width=0.49\textwidth] {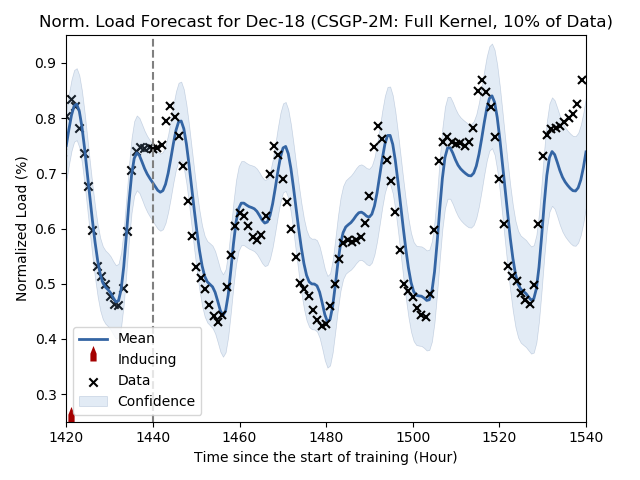}
		} \\
	\subfloat[Price forecast with 10\% of 3M data]{
		\includegraphics[width=0.49\textwidth] {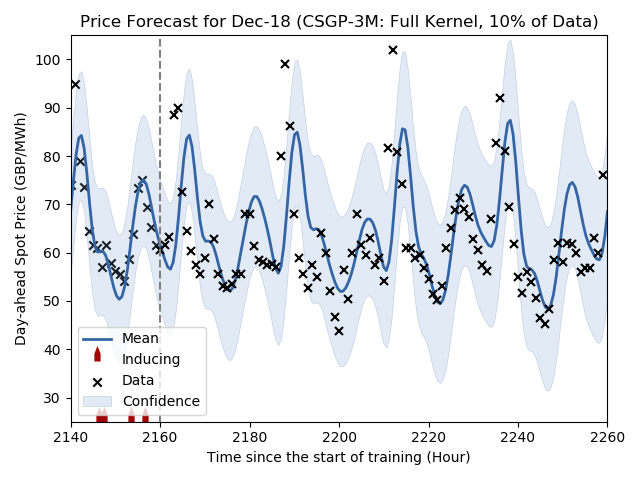}
		\label{fig:csgp_3m_price}
		} 
	\hspace*{-1em}
	\subfloat[Load forecast with 10\% of 3M data]{
		\includegraphics[width=0.49\textwidth] {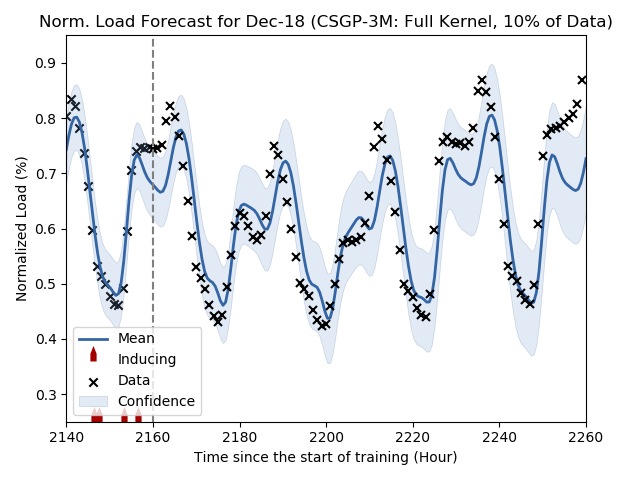}
		} \\
	\caption{Examining the effects of training on different lengths of data while keeping all other features fixed. The top pair trains on two months of hourly data while the bottom pair trains on three. Both are trained with 10\% sparsity.
	}
	\label{fig:exp_2}
\end{figure}

Fig. \ref{fig:exp_2} compares models trained on longer periods of two and three months. Upon inspection, it appears that the CSGP-1M with 10\% sparsity (the primary model) produces posterior predictions that are superior to the CSGP-2M and CSGP-3M that are both trained on 10\% sparsity. This is likely due to the recurrent nature of the data, so introducing more data from the past has only marginal benefits. Additionally, the repetitive nature of the data means that its truly unique segment is actually a subset that is ‘replicated’ many times. This has the effect of cramming more points (144 and 216 for CSGP-2M and CSGP-3M respectively) into the unique segment of the data, which shortens the correlation length across points. The model consequently learns less of the broader trend but more of the random noise (See Section \ref{sec_gp_comp_exp_1}). For instance, the posterior means in both Fig. \ref{fig:csgp_2m_price} and Fig. \ref{fig:csgp_3m_price} have missed the cluster of points at around the 1480$^{\textrm{th}}$ and 2200$^{\textrm{th}}$ hour respectively. As shown in Fig. \ref{fig:csgp_1m_sp010_price}, the primary model does not have this problem.

\section{Empirical hedging with the GP model}
\label{sec:hedging_results}

\subsection{Hedging results}

For the average load model, we assume an oracle that is able to look forward in time to know the exact load for the hedging month. To this end, $V^{b*}$ is obtained by taking the mean of actual hourly load during off-peak hours. $V^{p*}$ is obtained in a similar manner with a slight modification: first by taking the mean of hourly load for peak hours and then subtracting $V^{b*}$ from the result. Assuming spot-price and consumption load is uncorrelated, hedging at the expected load results in the minimum variance position; this is a robust approach used by industry (\citet{tegner_risk-minimisation_2017}). While it is entirely plausible that non-public, proprietary models exist that perform better than this benchmark, we do not concern ourselves with what those might be.  

\begin{figure}[th!]
	\centering
	\subfloat[Absolute payoffs across models]{
		\includegraphics[width=0.49\textwidth] {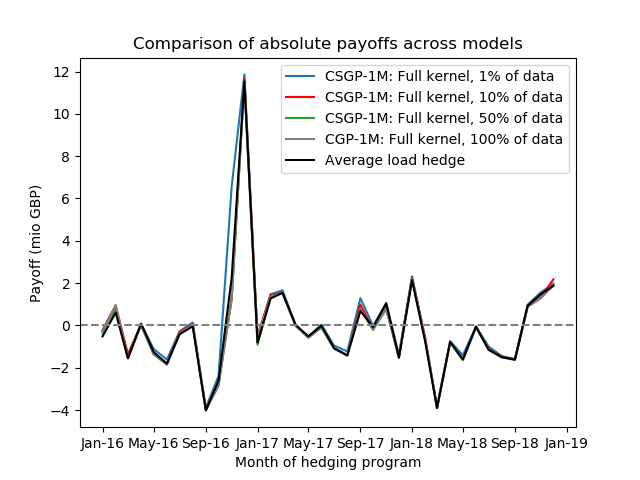}
		\label{fig:abs_payoff_3a}
		} 
	\hspace*{-1em}
	\subfloat[Cumulative payoffs across models]{
		\includegraphics[width=0.49\textwidth] {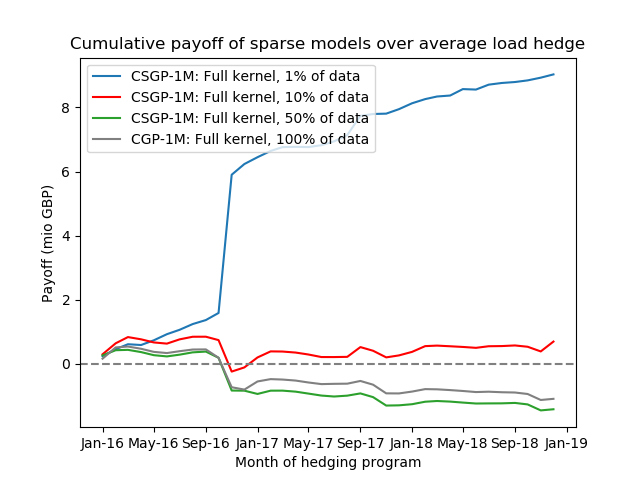}
		\label{fig:cum_payoff_3b}
		} \\
	\caption{Comparing hedging performance of various GP models against the average load hedge. Left compares absolute monthly payoff, right shows the cumulative payoff in excess of the comparator. Both charts are plotted over the span of the empirical hedging experiment (January 2016 to December 2018).}
	\label{fig:hedge_perf}
\end{figure}

The coregionalized sparse GP models outperforms the average load comparator across the duration of the hedging program. Appendix \ref{appen:B} provides a monthly breakdown with the corresponding optimal $V^b$ and $V^p$ (scaled by the average load for the comparator, and the maximum load across the data for the model). Fig. \ref{fig:hedge_perf} illustrates the result – the chart on the left compares absolute monthly performance while the right shows the cumulative performance over the average load hedge. Both charts are plotted over the duration of the experiment.

\begin{figure}[th!]
	\centering
	\subfloat[Load forecast made with 1\% of data]{
		\includegraphics[width=0.49\textwidth] {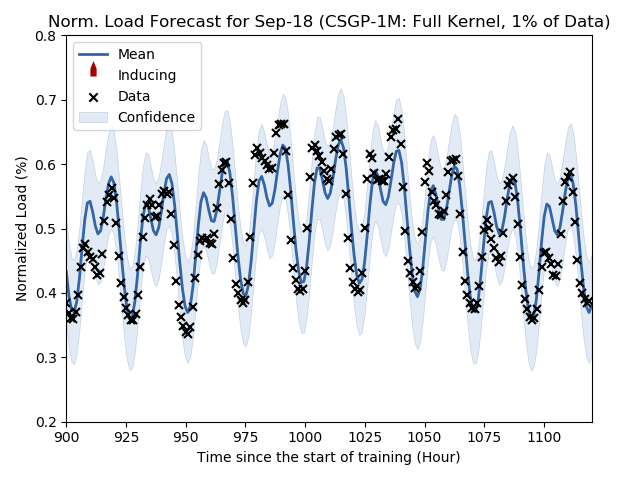}
		} 
	\hspace*{-1em}
	\subfloat[Load forecast made with 100\% of data]{
		\includegraphics[width=0.49\textwidth] {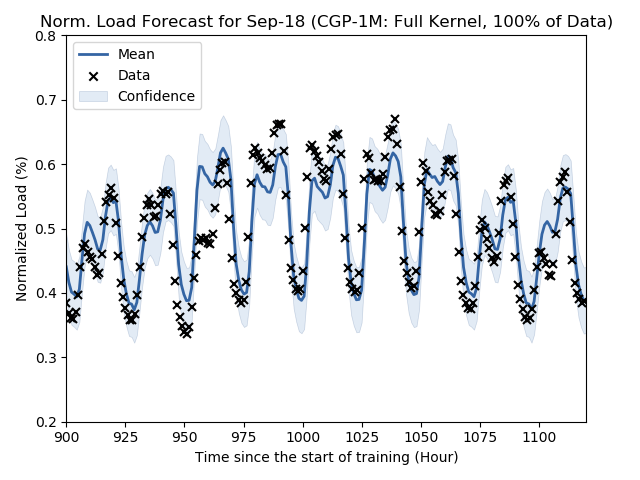}
		\label{fig:sparsity_comparison_4b}
		} \\
	\caption{Comparing the posterior on normalized load across different levels of sparsity. The model on the left uses 1\% of the data, the right uses the entire training set.}
	\label{fig:sparsity_comparison_50}
\end{figure}

On a monthly basis, the CSGP-1M with 10\% sparsity (which is our primary model) outperforms the average load hedge on the majority of the months over the three year period from January 2016 to December 2018. The relative performances of the various GP models in Fig. \ref{fig:abs_payoff_3a} are difficult to discern, but the cumulative plot on the right makes things clear.
From Fig. \ref{fig:cum_payoff_3b}, it appears that hedging performance is correlated with sparsity since it is the CSGP-1M with 1\% sparsity that delivers the highest payoff in excess of the average load. Over the 3 years spanning 2016-2018, the CSGP-1M with 1\% and 10\% sparsity respectively has payoffs of 5.42 and -2.91 mio GBP, both higher than the average load's -3.61 mio GBP (See the 'Total payoffs' row in Appendix \ref{appen:B}). 

As explained in Section \ref{sec_gp_comp_exp_1}, decreasing the number of data points increases the length scale; this focuses the learning on longer term trends rather than the noise in the data. We use the forecasts made by the CSGP-1M with 1\% sparsity (using 1\% of data) and CGP-1M (using the full data set) for September 2018 as an example. The results are shown in Fig. \ref{fig:sparsity_comparison_50}. Both forecasts appear to be reasonably similar, although we argue that the CSGP-1M is superior since its posterior mean is closer to the actual data points. Furthermore, most if not all points are well-enclosed in its confidence interval.
The CGP-1M uses all 720 hours for training. This produces a shorter length scale, which ends up fitting noise. This has the effect of generating overly confident predictions, as can be seen by the cluster of points around the 950$^\textrm{th}$ hour and the upper cluster of points in the 975-1040 hour range not contained within the confidence interval in Fig. \ref{fig:sparsity_comparison_4b}.

\subsection{Caveats and weaknesses}
There are some caveats to keep in mind. Firstly, we recognize the myriad of intricacies pertaining to the actual hedging of electricity. However, we had to simplify certain aspects due to the lack of public data and to also balance between real life applicability and keeping to a reasonable scope of work. Some examples are our approach of deriving consumption load from demand, as well as the pre-processing step of smoothing out price spikes by capping them to be at most three standard deviations away from the mean. As we had no access to the typical level of load that retailers aim to meet, another crucial set of simplifications we make in order to calculate payoffs is the following: (i) we adopt the position of a firm supplying 1.5\% of the market (Assuming around 70 domestic suppliers (\citet{ofgem_active_suppliers_2018}) as well as equal market share across participants, this is approximately a share of about 1.42\% per retailer. We round this figure up to 1.5\% for convenience), (ii) the area that this retailer is supplying has a similar load profile to the national-level load profile, and thus (iii) the profile that this retailer is supply to is a constant 1.5\% ‘strip’ of the national-level load.

\section{Conclusion}
\label{sec:conclusion}

We demonstrate that GPs can play a significant role in the risk management pipeline for the power markets. Additionally, coregionalization also shows that there is knowledge transfer between the GPs trained on spot price and load. Lastly, sparsity allows us to, with a smaller computational load, produce forecasts that are comparable to similar models that are trained on far much more data. We highlight some potentially interesting directions that future research we can take.

A straightforward extension would be to further pre-process the data in order to improve accuracy. For instance, we can further segment and fit additional separate models for different days (holidays, weekends) and then combining their respective predictions. It is likely that this approach will improve the fit for the GP model and hence hedging efficacy. 

Another possible extension would be to re-cast the problem as a dynamic hedging program. In other words, $V^b$ and $V^p$ are no longer fixed at the start of the month but are instead adjusted over the course of hedging period. This approach requires a dynamic model of the various price and load variables, as well as other modifications to the existing framework.

Regarding kernel design, our approach in this paper is not exhaustive. It is entirely likely that other sophisticated high-performing compositions exist. With the recent focus on AI explainability, there is a risk that these ‘black box’ kernels may be challenging to interpret. To this end, \citet{duvenaud_structure_2013} offers a method that might serve as a helpful starting point. 

Lastly, another avenue for further work is modifying the kernel parameter estimation task to give it a more Bayesian flavor. In this form, we move from estimating a posterior for the processes to also estimating posteriors for parameters. We can do this by first introducing some prior belief on what each parameter estimate should take. Subsequently and by updating based on observed data, we obtain a joint posterior distribution for each estimate. Our final maximum likelihood 'guess' of the true value for a parameter’s estimate can be obtained by summing over the posterior. 

\printbibliography

\clearpage 

\appendix

\begin{table}[!htb]
\section{Dates at which the hedging for the month was initiated}
\label{appen:A}
\centering
    \newcommand{\ra}[1]{\renewcommand{\arraystretch}{#1}}
    \begin{minipage}[t]{\linewidth}
      \centering
\ra{1.3}
\begin{tabular}{@{}lllcll@{}}\toprule
Contract & Init. Date & & &  Contract & Init. Date  \\
\midrule
Jan-16         & 18-Dec-15 & & & Jul-17         & 16-Jun-17     \\
Feb-16         & 18-Jan-16 & & & Aug-17         & 18-Jul-17     \\
Mar-16         & 16-Feb-16 & & & Sep-17         & 18-Aug-17     \\
Apr-16         & 21-Mar-16 & & & Oct-17         & 15-Sep-17     \\
May-16         & 18-Apr-16 & & & Nov-17         & 18-Oct-17     \\
Jun-16         & 19-May-16 & & & Dec-17         & 17-Nov-17     \\
Jul-16         & 20-Jun-16 & & & Jan-18         & 18-Dec-17     \\
Aug-16         & 18-Jul-16 & & & Mar-18         & 15-Feb-18     \\
Sep-16         & 18-Aug-16 & & & Apr-18         & 16-Mar-18     \\
Oct-16         & 19-Sep-16 & & & May-18         & 17-Apr-18     \\
Nov-16         & 18-Oct-16 & & & May-18         & 17-Apr-18     \\
Dec-16         & 17-Nov-16 & & & Jun-18         & 18-May-18     \\
Jan-17         & 19-Dec-16 & & & Jul-18         & 18-Jun-18     \\
Feb-17         & 18-Jan-17 & & & Aug-18         & 18-Jul-18     \\
Mar-17         & 15-Feb-17 & & & Sep-18         & 17-Aug-18     \\
Apr-17         & 20-Mar-17 & & & Oct-18         & 17-Sep-18     \\
May-17         & 18-Apr-17 & & & Nov-18         & 18-Oct-18     \\
Jun-17         & 18-May-17 & & & Dec-18         & 16-Nov-18     \\
\bottomrule
\end{tabular}
    \caption*{To hedge for a particular month, we hold the appropriate amount of base and peak load contracts approximately 2 weeks before the start of that month given by the above dates. The base and peak load prices used for a particular month are the closing price for each respective forward contract.}
    \end{minipage}%
\end{table}

\clearpage

\begin{landscape}

\newcommand{\ra}[1]{\renewcommand{\arraystretch}{#1}}
\begin{table*}\centering

\section{Comparing results of the model hedge and the average load}
\label{appen:B}
\ra{1.3}
\begin{tabular}{@{}rrrrcrrrcrrr@{}}\toprule
& \multicolumn{3}{c}{Average Load Hedge} & \phantom{abc}& \multicolumn{3}{c}{CSGP-1M 1\% Sp.} & \phantom{abc} & \multicolumn{3}{c}{CSGP-1M 10\% Sp.}\\
\cmidrule{2-4} \cmidrule{6-8} \cmidrule{10-12}& $V^{b*}$ & $V^{p*}$ & Payoff && $V^{b*}$ & $V^{p*}$ & Payoff && $V^{b*}$ & $V^{p*}$ & Payoff\\
\midrule
Jan-16         & 0.61                                                           & 0.19 & -0.52                                                            &  & 0.59                                                               & 0.44 & -0.28  &  & 0.52 & 0.41 & -0.22  \\
Feb-16         & 0.66                                                           & 0.17 & 0.61                                                             &  & 0.46                                                               & 0.28 & 0.82   &  & 0.46 & 0.43 & 0.95   \\
Mar-16         & 0.64                                                           & 0.13 & -1.56                                                            &  & 0.70                                                               & 0.46 & -1.40  &  & 0.74 & 0.28 & -1.37  \\
Apr-16         & 0.72                                                           & 0.13 & 0.09                                                             &  & 0.64                                                               & 0.22 & 0.06   &  & 0.84 & 0.01 & 0.02   \\
May-16         & 0.69                                                           & 0.13 & -1.27                                                            &  & 0.62                                                               & 0.42 & -1.12  &  & 0.66 & 0.10 & -1.37  \\
Jun-16         & 0.69                                                           & 0.16 & -1.81                                                            &  & 0.62                                                               & 0.52 & -1.61  &  & 0.60 & 0.12 & -1.84  \\
Jul-16         & 0.70                                                           & 0.15 & -0.42                                                            &  & 0.37                                                               & 0.23 & -0.28  &  & 0.39 & 0.21 & -0.29  \\
Aug-16         & 0.67                                                           & 0.13 & -0.04                                                            &  & 0.46                                                               & 0.31 & 0.14   &  & 0.47 & 0.18 & 0.04   \\
Sep-16         & 0.65                                                           & 0.16 & -4.02                                                            &  & 0.42                                                               & 0.28 & -3.89  &  & 0.43 & 0.22 & -4.01  \\
Oct-16         & 0.63                                                           & 0.16 & -2.60                                                            &  & 0.37                                                               & 0.10 & -2.38  &  & 0.49 & 0.04 & -2.71  \\
Nov-16         & 0.63                                                           & 0.19 & 2.24                                                             &  & 0.24                                                               & 0.12 & 6.56   &  & 0.35 & 0.06 & 1.27   \\
Dec-16         & 0.60                                                           & 0.18 & 11.53                                                            &  & 0.12                                                               & 0.19 & 11.89  &  & 0.59 & 0.14 & 11.66  \\
Jan-17         & 0.65                                                           & 0.20 & -0.81                                                            &  & 0.58                                                               & 0.27 & -0.60  &  & 0.62 & 0.28 & -0.50  \\
Feb-17         & 0.63                                                           & 0.18 & 1.27                                                             &  & 0.49                                                               & 0.22 & 1.47   &  & 0.49 & 0.34 & 1.46   \\
Mar-17         & 0.62                                                           & 0.13 & 1.54                                                             &  & 0.78                                                               & 0.35 & 1.67   &  & 0.79 & 0.46 & 1.54   \\
Apr-17         & 0.70                                                           & 0.10 & 0.02                                                             &  & 0.65                                                               & 0.19 & 0.03   &  & 0.76 & 0.01 & -0.01  \\
May-17         & 0.69                                                           & 0.12 & -0.52                                                            &  & 0.57                                                               & 0.19 & -0.52  &  & 0.67 & 0.01 & -0.58  \\
Jun-17         & 0.68                                                           & 0.14 & -0.02                                                            &  & 0.57                                                               & 0.17 & 0.04   &  & 0.64 & 0.03 & -0.10  \\
\bottomrule
\end{tabular}
\caption*{Optimal holdings of base and peak load contracts (\%) and the corresponding payoffs (mio GBP) for the Average Load hedge as well as the CSGP-1M with 1\% Sparsity and CSGP-1M with 10\% Sparsity. The CSGP-1M with 1\% Sparsity is the only model that ends the hedging program with a positive payoff.}
\end{table*}

\end{landscape}

\clearpage

\begin{landscape}

\newcommand{\ra}[1]{\renewcommand{\arraystretch}{#1}}
\begin{table*}\centering
\section*{Comparing results of the model hedge and the average load (continued)}
\ra{1.3}
\begin{tabular}{@{}rrrrcrrrcrrr@{}}\toprule
& \multicolumn{3}{c}{Average Load Hedge} & \phantom{abc}& \multicolumn{3}{c}{CSGP-1M 1\% Sp.} & \phantom{abc} & \multicolumn{3}{c}{CSGP-1M 10\% Sp.}\\
\cmidrule{2-4} \cmidrule{6-8} \cmidrule{10-12}& $V^{b*}$ & $V^{p*}$ & Payoff && $V^{b*}$ & $V^{p*}$ & Payoff && $V^{b*}$ & $V^{p*}$ & Payoff\\
\midrule
Jul-17         & 0.68                                                           & 0.13 & -1.08                                                            &  & 0.43                                                               & 0.20 & -0.96  &  & 0.47 & 0.11 & -1.08  \\
Aug-17         & 0.68                                                           & 0.13 & -1.42                                                            &  & 0.50                                                               & 0.38 & -1.23  &  & 0.54 & 0.11 & -1.42  \\
Sep-17         & 0.65                                                           & 0.15 & 0.69                                                             &  & 0.01                                                               & 0.05 & 1.29   &  & 0.16 & 0.19 & 0.99   \\
Oct-17         & 0.57                                                           & 0.16 & -0.09                                                            &  & 0.39                                                               & 0.15 & -0.02  &  & 0.49 & 0.06 & -0.20  \\
Nov-17         & 0.62                                                           & 0.19 & 1.04                                                             &  & 0.33                                                               & 0.12 & 1.06   &  & 0.52 & 0.07 & 0.84   \\
Dec-17         & 0.62                                                           & 0.18 & -1.52                                                            &  & 0.43                                                               & 0.20 & -1.39  &  & 0.40 & 0.32 & -1.46  \\
Jan-18         & 0.64                                                           & 0.20 & 2.14                                                             &  & 0.46                                                               & 0.31 & 2.32   &  & 0.58 & 0.27 & 2.25   \\
Feb-18         & 0.67                                                           & 0.16 & -0.76                                                            &  & 0.59                                                               & 0.29 & -0.63  &  & 0.59 & 0.33 & -0.58  \\
Mar-18         & 0.63                                                           & 0.14 & -3.91                                                            &  & 0.60                                                               & 0.31 & -3.83  &  & 0.71 & 0.18 & -3.89  \\
Apr-18         & 0.68                                                           & 0.13 & -0.78                                                            &  & 0.80                                                               & 0.99 & -0.75  &  & 0.72 & 0.04 & -0.80  \\
May-18         & 0.70                                                           & 0.10 & -1.61                                                            &  & 0.66                                                               & 0.50 & -1.41  &  & 0.70 & 0.09 & -1.63  \\
Jun-18         & 0.71                                                           & 0.12 & -0.05                                                            &  & 0.43                                                               & 0.05 & -0.07  &  & 0.48 & 0.05 & -0.08  \\
Jul-18         & 0.74                                                           & 0.12 & -1.15                                                            &  & 0.45                                                               & 0.29 & -1.00  &  & 0.46 & 0.13 & -1.10  \\
Aug-18         & 0.70                                                           & 0.13 & -1.50                                                            &  & 0.43                                                               & 0.16 & -1.45  &  & 0.45 & 0.10 & -1.49  \\
Sep-18         & 0.64                                                           & 0.13 & -1.62                                                            &  & 0.36                                                               & 0.11 & -1.59  &  & 0.37 & 0.11 & -1.60 \\
Oct-18         & 0.64                                                           & 0.13 & 0.93                                                            &  & 0.36                                                               & 0.11 & 0.98  &  & 0.37 & 0.11 & 0.89 \\
Nov-18         & 0.64                                                           & 0.13 & 1.48                                                            &  & 0.36                                                               & 0.11 & 1.56  &  & 0.37 & 0.11 & 1.33 \\
Dec-18         & 0.64                                                           & 0.13 & 1.87                                                            &  & 0.36                                                               & 0.11 & 1.98  &  & 0.37 & 0.11 & 2.18 \\
\bottomrule
\textbf{Total Payoff}         &  &  & \textbf{-3.61}                                                            &  & & & \textbf{5.42}  & & & & \textbf{-2.91}
\\
\midrule
\end{tabular}
\caption*{Optimal holdings of base and peak load contracts (\%) and the corresponding payoffs (mio GBP) for the Average Load hedge as well as the CSGP-1M with 1\% Sparsity and CSGP-1M with 10\% Sparsity. The CSGP-1M with 1\% Sparsity is the only model that ends the hedging program with a positive payoff.}
\end{table*}

\end{landscape}




\end{document}